\documentclass[12pt,preprint]{aastex}

\newcommand {\kms} {\,{\rm km\,s}^{-1}}

\usepackage{rotating}

\shorttitle{BSSs in NGC 6752}
\shortauthors{Lovisi et al.}
 
\begin{document} 
\title{Another brick in understanding chemical and kinematical properties of BSSs: NGC 6752
\footnote{Based on FLAMES observations collected at the European Southern Observatory, proposal numbers 
081.D-0356 and 089.D-0298.}}
\author{
L. Lovisi\altaffilmark{1},
A. Mucciarelli\altaffilmark{1}, 
E. Dalessandro\altaffilmark{1},
F.R. Ferraro\altaffilmark{1}, 
B. Lanzoni\altaffilmark{1}
\affil{\altaffilmark{1} Dipartimento di Fisica \& Astronomia, Universit\`a degli Studi
di Bologna, viale Berti Pichat 6/2, I--40127 Bologna, Italy}%
} 
\date{}
 
\begin{abstract} 
We used high-resolution spectra acquired with the
multifiber facility FLAMES at the Very Large Telescope of the European
Southern Observatory to investigate the chemical and kinematical
properties of a sample of 22 Blue Straggler Stars (BSSs) and 26 red giant
branch stars in the nearby globular cluster NGC 6752. We measured
radial and rotational velocities and Fe, O and C abundances. According to radial 
velocities, metallicity and proper motions we identified 18 BSSs as likely cluster members.
We found that all the BSSs rotate slowly (less than 40$\kms$), similar to the findings
in 47 Tucanae, NGC 6397 and M30. The Fe abundance analysis reveals the 
presence of 3 BSSs affected by radiative levitation (showing [Fe/H] significantly higher than that measured in 
``normal`` cluster stars), confirming that 
element transport mechanisms occur in the photosphere of BSSs hotter than $\simeq$8000 K.
Finally, BSS C and O abundances are consistent with those measured in dwarf stars. No C and O
depletion ascribable to mass transfer processes has been found on the atmospheres of the studied BSSs 
(at odds with previous results for 47 Tucanae and M30), suggesting
the collisional origin for BSSs in NGC 6752 or that the CO-depletion is a transient phenomenon.

\end{abstract} 
 

\section{Introduction}
\label{intro}
Galactic globular clusters (GCs) are dynamically active stellar systems where interactions
and collisions, especially involving binaries, are quite frequent \citep{Hut92, Meylan97}.
These interactions likely generate exotic objects like X-ray binaries, millisecond pulsars and 
Blue Straggler stars (BSSs; see \citealt{Paresce92, Baylin95, Bellazzini95, Ferraro95, Ferraro09,
Ransom05, Pooley06}). BSSs are the most abundant products of this activity.
They are brighter and bluer than the main sequence (MS) turnoff (TO).
They lie along an extension of the MS and they are known to be more massive than normal TO stars
\citep{Shara97, Gilliland98, DeMarco05}. Because of their mass, BSSs are invaluable probes of 
GC internal dynamics. In particular, it has recently shown that
their radial distribution is a powerful tool to measure the cluster 
dynamical age \citep{Ferraro12}. Nevertheless, many basic questions about the formation mechanisms and the physical properties
of these puzzling objects remain open. Two main scenarios have been proposed for their formation: they could form through mass transfer in a
binary system that could lead to the complete coalescence of the two companions (MT-BSSs; \citealt{McCrea64}) or
through stellar collisions (COL-BSSs; \citealt{HillsDay76}).
Hydrodynamic simulations suggest different chemical patterns for BSSs formed through the two different mechanisms.
In particular, MT-BSSs are expected to show C and O depletion \citep{Sarna96} whereas no chemical anomalies are predicted for 
COL-BSSs \citep{Lombardi95}. On the contrary, predictions about the BSS rotational velocity v$_{e}$ $\sin(i)$
are quite controversial. In fact, both MT- and COL-BSSs are expected to rotate fast 
\citep{Benz87, Sarna96}, but accurate simulations are lacking for MT-BSSs and braking mechanisms may intervene both 
for MT- and COL-BSSs slowing down the initial BSS rotation, with timescales and efficiencies still unknown \citep{Leonard95, Sills05}. 

In addition to GCs, BSSs have been detected in a variety of low-density environments such as Galactic open clusters \citep{Ahumada95}, 
dwarf spheroidal galaxies in the Local Group \citep{Momany07}, the Milky Way's bulge \citep{Clarkson11} and halo
\citep{Preston00, Sneden03, Carney05} and, recently, also ultra-faint galaxies \citep{Santana13}. 
Despite less populous than GCs, these low-density systems contain a high percentage of BSSs. However, the low collisional rate in
low-density environments implies a different origin for the BSSs in GCs and loose systems, where BSSs likely formed exclusively by MT.

In this framework, we started an observational campaign aimed at performing the first systematic study of the chemical and kinematical properties
of BSSs in GCs and at studying their possible differences in MT- and COL-BSSs. Spectroscopic results by \citet{Ferraro06} have observationally 
revealed the presence of a subpopulation of CO-depleted BSSs in 47 Tucanae and similarly \citet{Lovisi13} identified 4 BSSs with O upper limits 
incompatible with the cluster O abundance in M30. In both cases, the amount of the observed C and O depletion is incompatible with 
that expected for second generation stars (predicted to be C and O poor) and cannot be interpreted in terms of self-enrichment processes. 
On the contrary, \citet{Lovisi10} found no evidence of CO-depleted BSSs in M4, whereas 
no interpretation in terms of formation mechanism was possible for the BSSs studied by \citet{Lovisi12} in NGC 6397,
due to the occurrence of a diffusion process, called radiative levitation, that alters chemical abundances on the
surfaces of stars with shallow or no-convective envelopes. \citet{Ferraro06} identified 
only one fast rotating (v$_{e}$ $\sin(i)>$50 $\kms$) BSS in 47 Tucanae, corresponding to $\sim$2\% of the studied population. 
A similar fraction of fast rotating BSSs has been also found by \citet{Lovisi12} and \citet{Lovisi13} (respectively 
$\sim$5\% and $\sim$6\%). Surprisingly, \citet{Lovisi10} found that 40\% of the studied BSS population in M4 rotate faster than 50 $\kms$.

In this work, we discuss the properties of a BSS sample in NGC 6752. The paper is structured as follows:
the observations are described in Sect. \ref{obs}. The determination of stellar radial velocities and cluster membership are 
discussed in Sect. \ref{v_rad}. The estimate of atmospheric parameters is described in Sect. \ref{par}. The measured rotational 
velocities are presented in Sect. \ref{rot}. Sect. \ref{chem} describes methods and results of the chemical abundance
analysis for our sample. Finally our conclusions are drawn in Sect. \ref{discuss}.

\section{Observations}
\label{obs}
This work is based on the analysis of spectra of individual stars in
NGC 6752, obtained with the multi-object facility FLAMES \citep{Pasquini02}
mounted at the Unit Telescope 2 at the Very Large Telescope of the European 
Southern Observatory. Observations were performed
during two different runs (081.D-0356 and 089.D-0298, hereafter
P81 and P89, respectively) with the UVES+GIRAFFE combined mode, during 
the nights 2008 August 13-16 and 2012 July 4.
Spectra have been acquired for 22 BSSs and 26 red giant branch (RGB) stars, 
most of them being observed in both P81 and P89. 
Four different setups have been used for the spectroscopic observations:
HR5A (with spectral coverage 4340-4587 \AA\ and spectral resolution R=18470), 
HR15N (6470-6790 \AA, R=17000), HR18 (7468-7889 \AA, R=18400) and 
HR22A (8816-9565 \AA, R=11642), suitable to sample some metallic lines,
the H$\alpha$ line, the O~I triplet at $\simeq 7774$ \AA\ and two C~I lines at
9078.3 and 9111.8 \AA, respectively. Multiple exposures for each target 
have been secured according to the efficiency of the different setups, in particular
5 exposures with HR5A, 2 with HR15N and 4 with HR18 and HR22A have been obtained.
The total exposure times amount to 3.75 hours for the HR5A, 1.5 
hours for the HR15N and 3 hours for the HR18 and HR22A. 
The spectroscopic target selection has been performed from a photometric catalog obtained by 
combining ACS@HST data in V and I bands for the central regions and WFI@ESO observations 
in B and R bands for the external regions. Fig. \ref{cmd} shows the CMDs for the ACS and WFI datasets, with the position 
of all the analyzed targets. The spectra pre-reduction has been performed by using the latest 
version of the GIRAFFE ESO pipeline (\textit{gasgano} version 2.4.3\footnote{http://www.eso.org/sci/software/pipelines/}), 
that includes bias subtraction, flat-field correction, wavelength calibration and 
one-dimensional spectra extraction. For each exposure we subtracted the corresponding 
master sky spectrum, obtained by averaging several spectra of sky regions.
By combining the sky-subtracted exposures, we obtained median spectra with signal-to-noise 
ratios S/N$\simeq 30-120$ for the selected BSSs, and S/N$\simeq 50-150$ for the RGB stars.

\section{Radial velocities}
\label{v_rad}
\subsection{Cluster membership}
\label{rad}
\begin{figure}
\plotone{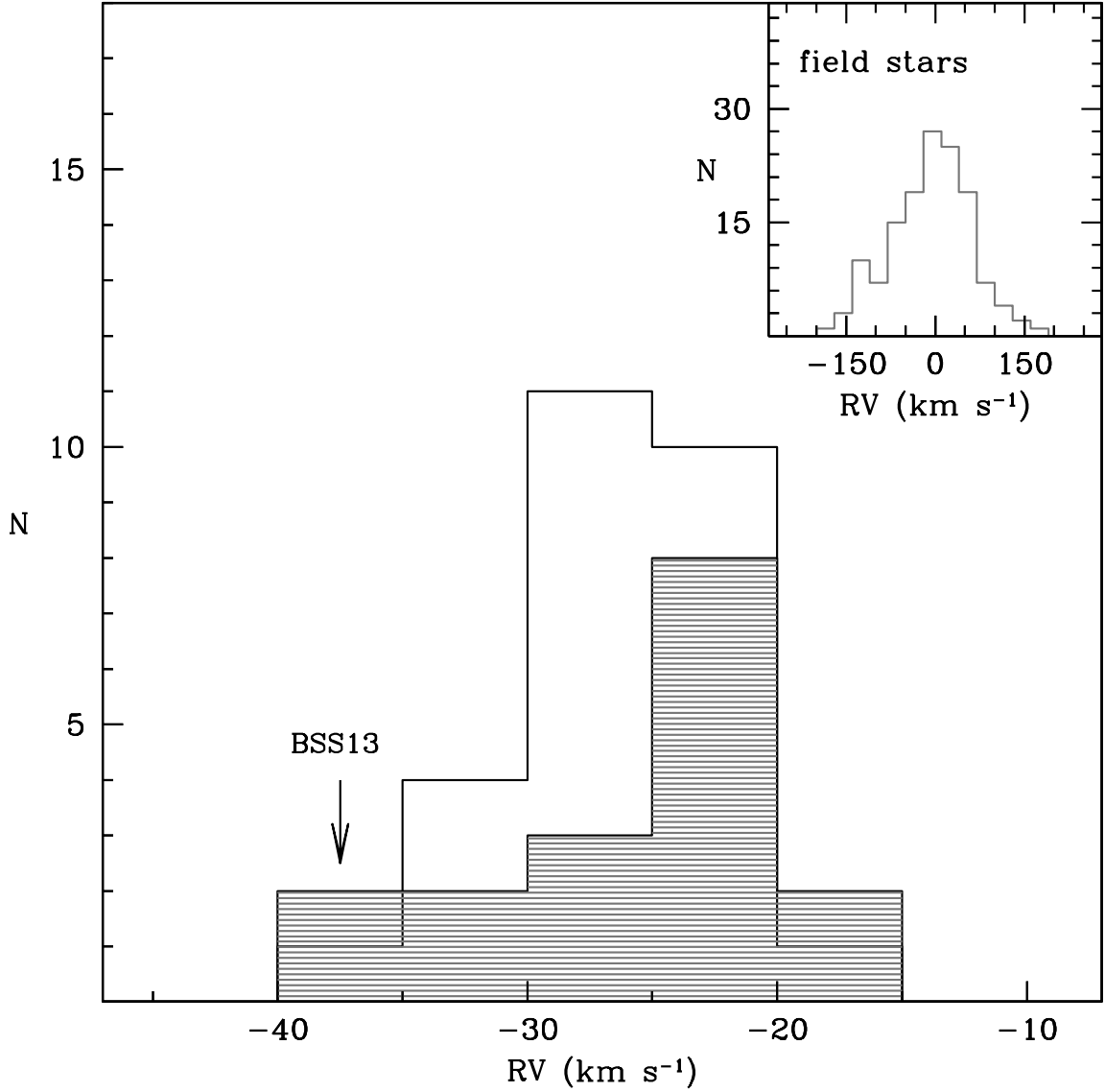}
\caption{Radial velocity distributions for the RGB stars (empty histogram) and the likely member BSSs (grey shaded histogram). The inset shows the radial
velocity distribution of field stars according to models by \citet{Robin03}.}
\label{vrad}
\end{figure}

In order to assess cluster membership for each target, radial velocities (RVs) have been measured
with the IRAF task \textit{fxcor} following the technique described by \citet{TonryDavis1979}. Synthetic 
spectra calculated with the atmospheric parameters of the analyzed targets (see Sect. \ref{par}) and convolved 
with a Gaussian profile to reproduce the spectral resolution of the corresponding GIRAFFE setup, 
have been used as templates for the cross-correlation (see Sect. \ref{rot} for details about 
the computation of the synthetic spectra). Finally, the heliocentric correction has been applied to all the RVs. 
Tables \ref{bss} and \ref{rgb} show the mean heliocentric RVs obtained as the average of values in P81
and P89.\\
The mean radial velocity of the RGB stars is RV=$-26.7\pm$0.8$\kms$ ($\sigma$=4.0$\kms$)  
which is fully in agreement with previous results by \citet{Harris96} and \citet{Carretta09}.
The derived RV distribution for the RGB stars is shown in Fig. \ref{vrad} (main panel, empty histogram). 
The mean RV of the RGB stars has been assumed as the systemic velocity of NGC 6752 and used to infer cluster membership
for BSSs: stars having RVs within 3$\sigma$ from the systemic velocity value have been 
considered as members of the cluster. The mean radial velocity of the 22 observed BSSs is 
RV=$-16.9\pm$0.8$\kms$ ($\sigma$=40.5$\kms$) but the 3$\sigma$-rejection based on the results of the RGB
stars leads to the identification of 4 BSSs (with RV=$-$92.8, $-$14.1, 97.2, 110.7 $\kms$) as possible non-members. 
However, particular discussion deserves BSS13. This star has RV=$-$14.1$\kms$ so that it is within 3.5$\sigma$ from the mean RV and
it is located very close (23.8\arcsec) to the cluster centre. Moreover, its
iron content is fully in agreement with the mean cluster metallicity (see Sect. \ref{chem}).
By using Besan\c{c}on Galactic model by \citet{Robin03}\footnote{http://model.obs-besancon.fr/}, 
we derived the metallicity distribution of field stars in the direction of NGC 6752, within the same 
magnitude and colour intervals shared by our BSS sample. We computed that the probability to 
observe a field star with an iron content similar to that of BSS13 ([Fe/H]$=-$1.43) is 1.4\%. 
For this reason, we decided to consider BSS13 as cluster member and to include it in our sample.

\begin{figure}
\plotone{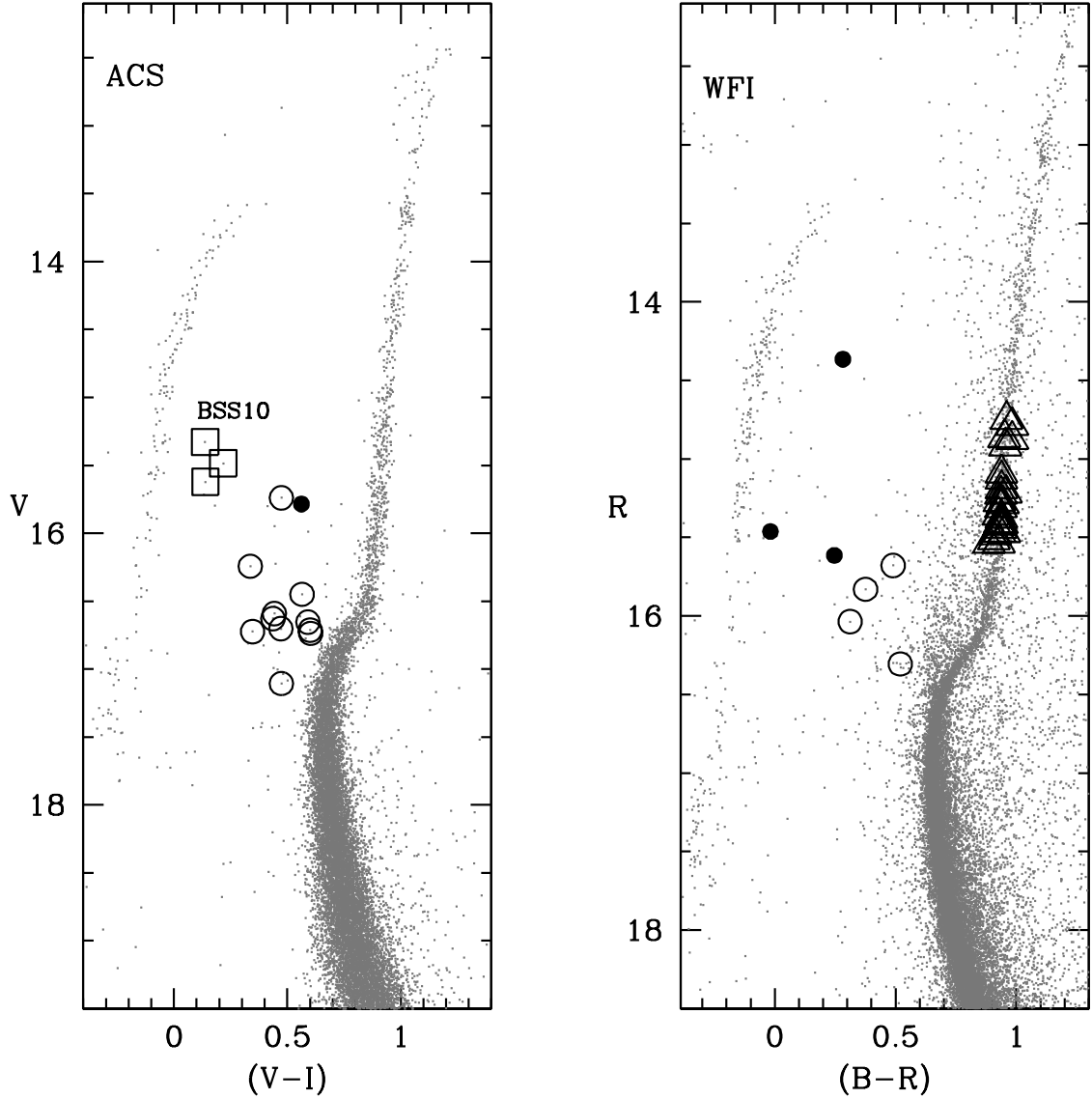}
\caption{Colour-magnitude diagrams of NGC 6752 for the ACS and WFI datasets. Open circles, squares and large black dots are the observed BSSs;
open triangles are the observed RGB stars. Large black dots are BSSs turned out to be field stars whereas open squares are BSSs affected 
by radiative levitation.}
\label{cmd}
\end{figure}

As shown in the inset of Fig. \ref{vrad} the systemic velocity of NGC 6752 is very similar 
to the peak of the RV distribution of field stars in the direction of the cluster
\citep{Robin03}, so that the probability of field 
contamination among the BSSs is very high and the measure of radial velocities 
cannot totally guarantee the cluster membership. Therefore, in order to further constrain the membership of BSSs in our sample, 
we used proper motions: we performed a cross-correlation between our BSS sample and the catalog of \citet{Zloczewski12} 
by using the software CataXcorr (P.Montegriffo, private communication). We found 9 BSSs in common. 
According to their proper motions, one BSS in our sample (with RV=$-$22.7 $\kms$, totally 
in agreement with the systemic velocity) certainly belongs to the field, one (namely BSS5) is a possible cluster member, 
whereas the remaining 7 stars (namely BSS1, BSS10, BSS11, BSS14, BSS15, BSS16 and BSS17) are likely members.
After removing non-member BSSs (large black dots in Fig. \ref{cmd}) from our sample, we are 
left with 18 BSSs that are likely cluster members (open circles in Fig. \ref{cmd}). The RV distribution for these stars is shown in Fig.
\ref{vrad} (grey shaded histogram). The mean value is RV=$-25.5\pm$0.9$\kms$ ($\sigma$=5.9$\kms$), in good 
agreement with the mean value inferred from the RGB stars.

In our sample we found 8 variable stars. In particular, BSS6, BSS10, BSS11 and BSS17 correspond to
V21, V17, V12 and V7 studied by \citet{Kaluzny09}. In particular, BSS11 and BSS17 are classified 
as SX-Phoenicis whereas BSS6 is likely a $\gamma$ Dor type pulsator and BSS10 has been classified 
as a possible binary composed of a hot extreme-Horizontal Branch star (EHB) and a red companion on 
the basis of its location in the CMD.\\
We have measured RV variation of  $\sim$ 13 $\kms$ and $\sim$ 5 $\kms$ 
between P81 and P89 for BSS11 and BSS17, both classified as SX-Phoenicis. 
Moreover, for BSS11 we observe differences in the RV also between individual HR5A exposures
(up to $\sim 25 \kms$). Given the duration of our exposures (45 minutes each), the variation of RV is in agreement
with the period found by \citet{Kaluzny09} for those SX-Phoenicis ($\sim 85$ and $\sim 59$ minutes).
However, we stress that the time sampling of our data set is not suitable to perform an accurate investigation of the radial 
velocity variability that can reveal possible binary stars. This does not rule out the possibility that some of the BSSs in our
sample belong to binary systems. 

\subsection{The nature of BSS10}
\begin{figure}
\plotone{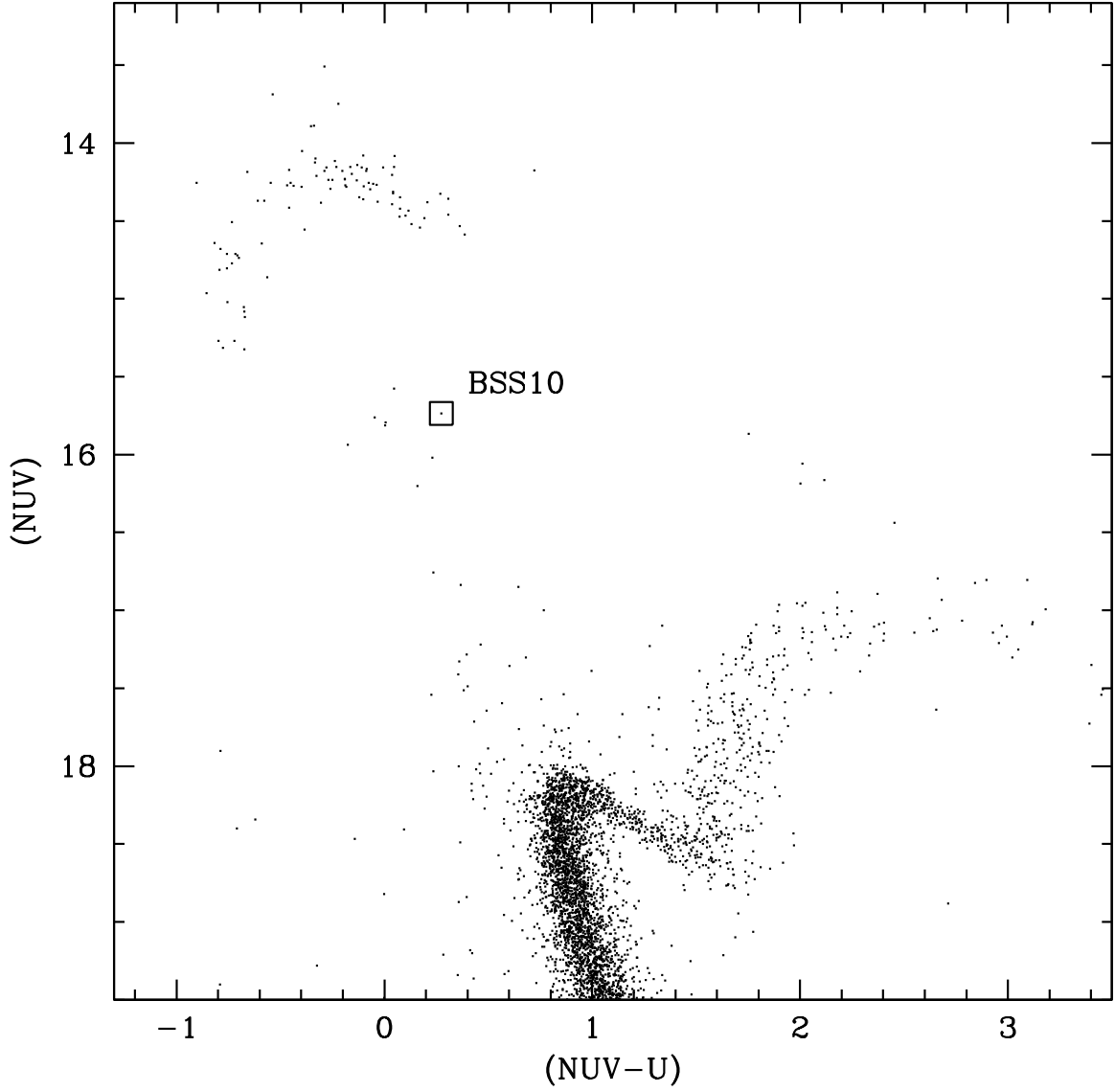}
\caption{CMD in the UV plane obtained from WFPC2 observations of NGC 6752 \citep{Sabbi04}. The open square marks the position of BSS10.}
\label{wfpc2} 
\end{figure}

A further clarification should be done also for BSS10. Based on its position in the CMD, 
\citet{Kaluzny09} suggested that this star could be an unresolved binary system formed by an EHB star and a red companion. 
As suggested by \citet{Ferraro97, Ferraro04}, the optimal way to observe hot stars like BSSs and Horizontal Branch stars 
is at ultraviolet (UV) wavelengths. In fact, in this plane RGB stars (the brightest objects in the optical CMD) are faint 
and the identification of hot stars is very easy. Therefore, in order to investigate the possibility that BSS10 could be an unresolved
binary system, we identified it in the UV WFPC2 catalog used by \citet{Sabbi04}. 
According to its position in the UV plane (see Fig. \ref{wfpc2}) 
we can conclude that BSS10 is too faint to be an EHB, but it is instead one of the brightest BSSs in NGC 6752.
We suggest that the variability observed in the light curve by \citet{Kaluzny09} might be due to a BSS in a
binary system with a low-mass MS companion. 

\section{Atmospheric parameters}
\label{par}

Effective temperature (T$_{eff}$) and surface gravity (log g) for all the observed targets have been derived 
comparing their position in the CMD with theoretical isochrones (with different ages) from the Padova database
\footnote{http://stev.oapd.inaf.it/cgi-bin/cmd.} \citep{Bressan12}. Given the stellar position in the CMD, 
each observed target has been orthogonally projected on the closest theoretical model, and T$_{eff}$ and 
log g have been derived for each star. For the RGB stars, an isochrone of 12 Gyr, Z=0.0005 and $\alpha$-enhanced 
chemical mixture has been superimposed on the CMD of NGC 6752, assuming a distance modulus 
(M-m)$_{0}=13.18$ and E(B-V)=0.04 \citep{Ferraro99}. For BSSs, isochrones with different ages (2-8 Gyrs, and assuming the same 
chemical composition, distance modulus and reddening values) have been used to sample the BSS region.

The atmospheric parameters for the BSS and RGB samples are listed in Tables \ref{bss} and \ref{rgb}, respectively.
Conservative errors in temperatures and gravities have been considered: 100 K and 0.2 dex both for BSSs and RGB stars.
Concerning microturbulent velocity, the small number of metallic lines 
prevents us to derive spectroscopically this parameter. For this reason, we assumed 2.0$\kms$ and 1.5$\kms$ 
for BSS and RGB stars, respectively. We finally adopted a conservative error of 0.5 $\kms$ for all the targets. 

\section{Rotational velocities} 
\label{rot}

Rotational velocities have been derived for all the targets through a $\chi^{2}$ minimization between the observed
spectrum and a grid of synthetic spectra, computed for all stars by adopting the estimated
atmospheric parameters and different rotational values. All the synthetic spectra have been computed by using 
the Kurucz's code SYNTHE \citep{Kurucz93, Sbordone04}, including the atomic data for all the lines from the most 
updated version of the Kurucz line list by F.Castelli \footnote{http://wwwuser.oat.ts.astro.it/castelli/linelists.html}. 
Model atmospheres for the synthetic spectra have been calculated with ATLAS9 code \citep{Kurucz93, Sbordone04} under the 
assumptions of Local Thermodynamic Equilibrium (LTE) and plane-parallel geometry and adopting the new opacity distribution 
functions by \citet{Castelli03}, without the inclusion of the approximate overshooting \citep{Castelli97}. 
The most prominent features in the stellar spectra have been used to infer stellar rotations.
In particular, for the RGB stars v$_{e}$ $\sin(i)$ values have been inferred from the Ba~II at 4554.03 \AA\ whereas 
for BSSs from the Mg~II line at 4481 \AA\ and the O~I line at 7771.95 \AA.
Typical uncertainties in the v$_{e}$ $\sin(i)$ measurement are of the order of few $\kms$.

\begin{figure}
\plotone{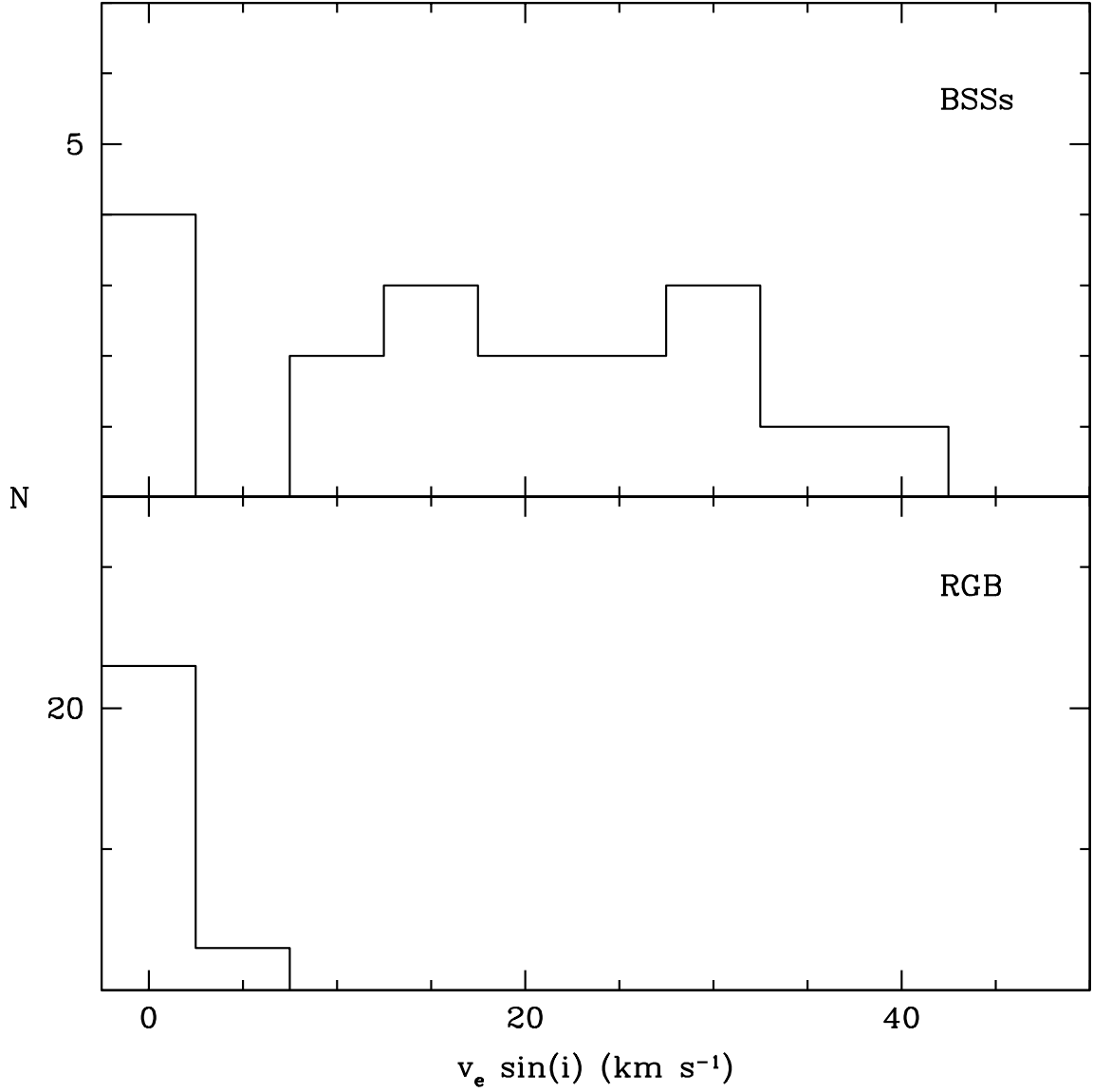}
\caption{Rotational velocity distributions for BSSs (upper panel) and RGB stars (lower panel).}
\label{rotbss}
\end{figure}

The rotational velocity distribution thus obtained for BSSs is shown in Fig. \ref{rotbss} (upper panel)
and it is compared with that of RGB stars (lower panel). 
Almost all RGB stars have v$_{e}$ $\sin(i) = 0 \kms$, the largest value not exceeding 6 $\kms$ (but with uncertainties that make them 
compatible with v$_{e}$ $\sin(i)\sim0-1 \kms$). 
These results are in agreement with those usually observed for this kind of stars
\citep[see e.g.][]{Carney08, Cortes09, Mucciarelli11, Lovisi13}. 
On the contrary, the rotational velocity distribution of BSSs is much wider: some BSSs show the same average v$_{e}$ $\sin(i)$
of RGB stars but others have rotational velocities several times larger, up to the highest value of $\sim$ 38 $\kms$.

Finally, Fig. \ref{fe} (upper panel) shows the rotational velocity as a function of temperature for all the targets: for
each temperature value high spread in v$_{e}$ $\sin(i)$ is observed.

\section{Chemical abundances}
\label{chem}

Iron abundances have been measured for all BSSs and compared with those derived from the RGB stars.
Oxygen abundances for BSSs have been derived from the O~I triplet at $\sim$7774 \AA\ and carbon 
abundances from two C~I lines at 9078.3 and 9111.8 \AA. No determination of O and C abundances is possible
for RGB stars of our sample, since their temperatures are too low to detect O and C lines
in the studied spectral ranges. Abundances for all the targets have been estimated with the code GALA 
\citep{Mucciarelli13a} by using the measured equivalent widths (EWs) of the absorption lines.
EWs for RGB stars have been measured with DAOSPEC \citep{Stetson08} whereas for BSSs we used
our own FORTRAN procedure \citep{Mucciarelli11} that has the advantage to perform a Gaussian fit on a spectral window selected interactively,
in order to optimize the line fitting procedure also for moderate rotating stars. 
The used reference solar abundances are from \citet{Grevesse98} for Fe and C and from 
\citet{Caffau11} for O.

Possible deviations from the LTE assumption have been taken into account. 
Non-LTE corrections for O abundances have been taken from \citet{Gratton99} and \citet{Takeda97} 
for BSSs colder and hotter than 8000 K respectively, and corrections from \citet{Tomkin92} have been used 
for C abundances. 
Conversely, no grid of non-LTE corrections are available in the literature for iron lines for the range of parameters 
typical of our BSSs. Based on the fact that non-LTE corrections for ionized elements in warm (A and F type) 
stars are negligible with respect to those for neutral elements, we obtained Fe abundances for the BSSs by 
using Fe~II lines, whereas Fe~I lines have been used for the RGB stars (for which the LTE approximation is 
still valid). Fe, C and O abundances for BSSs and RGB stars are listed in Table \ref{bss} and \ref{rgb}, respectively.
 
{\bf Iron --}
For the RGB stars we find an average iron abundance of [Fe/H]=$-$1.53$\pm$0.01 ($\sigma$=0.07) dex in very good agreement with 
results by \citet{Carretta09} and \citet{Yong13}. The mean iron abundance for BSSs is [Fe/H]=$-$1.41$\pm$0.01 ($\sigma$=0.29), 
that is 0.1 dex higher and more scattered than that of the RGB stars. Fig. \ref{fe} (lower panel) shows the comparison 
between the iron distributions of RGB stars (grey triangles) and BSSs (open circles and squares) as a function of temperatures: most of 
the BSSs show Fe values in agreement with the mean iron abundance for the RGB stars (grey dashed horizontal line), nevertheless the 3
hottest BSSs (T$_{eff}>$8000 K, open squares in Fig. \ref{fe}) show higher Fe abundances. The mean Fe abundance for BSSs colder than 8000 K
is [Fe/H]=$-$1.52$\pm$0.02 ($\sigma$=0.10), nicely matching the value found for the RGB stars.

\begin{figure}
\plotone{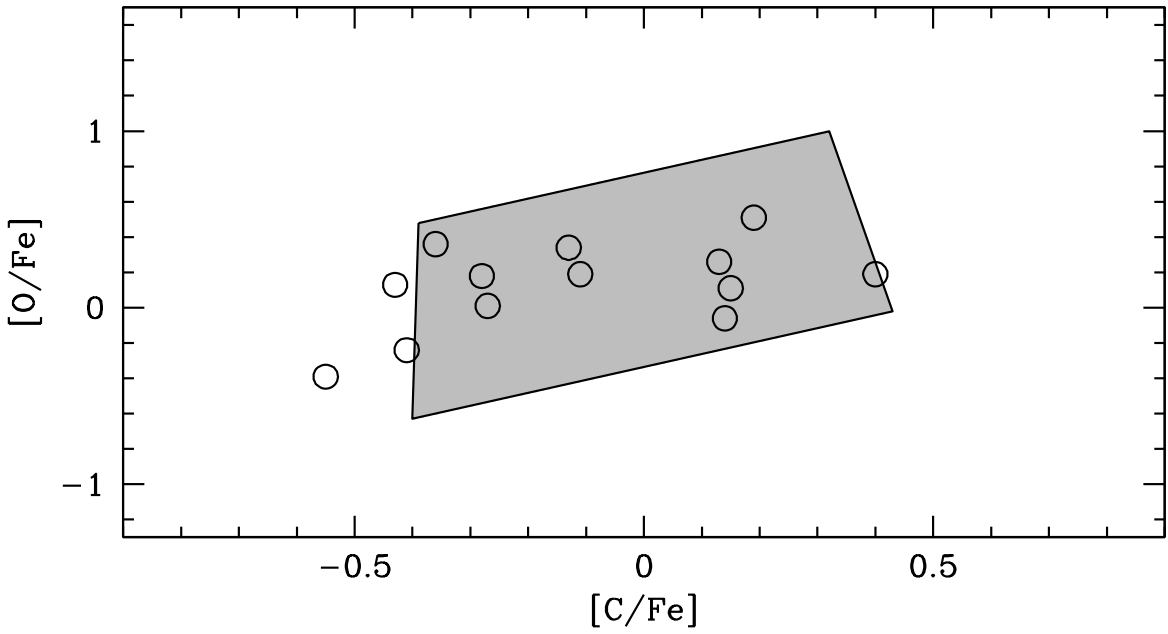}
\caption{[C/Fe] and [O/Fe] ratios of BSSs (open circles) compared with those of dwarf stars (grey region).}
\label{co}
\end{figure}

{\bf Carbon and oxygen --}
[C/Fe] and [O/Fe] ratios for BSSs not affected by radiative levitation are shown in Fig. \ref{co}. 
In order to consistently compare C and O abundances for BSSs with those of a parent population, we used results by 
\citet{Carretta05} obtained for dwarf stars (grey region in Fig. \ref{co}): BSS abundances well overlap 
the mean locus defined by the dwarf stars. Carbon and oxygen content for all
BSSs colder than 8000 K is [C/Fe]=$-$0.12$\pm$0.08 ($\sigma$=0.29) and [O/Fe]=$+$0.12$\pm$0.07 ($\sigma$=0.24).

\section{Discussion}
\label{discuss}

\subsection{Rotational velocities}
Rotational velocities have been derived for the BSSs and compared with values obtained for the RGB stars.
The BSS distribution is wider than the RGB one but no fast rotating (v$_{e}$ sin(i)$>$50$\kms$) 
BSSs have been found. In fact, apart from M4 (which is the most extreme case, \citealt{Lovisi10}) where $\sim$ 40\%
of BSSs have been found rotating faster than 50$\kms$, the other clusters studied so far (47 Tucanae \citealt{Ferraro06}, NGC 6397 \citealt{Lovisi12},
and M30 \citealt{Lovisi13}) show a small fraction of fast rotating BSSs ($\sim$2\%, $\sim$5\%, $\sim$6\% respectively,
that are also WUma or SX-Phoenicis stars). Assuming these low percentages, we should have found at most one fast rotating BSS. 
It is interesting to note that all the investigated clusters showing a small fraction of fast rotating BSSs are high density
clusters, possibly dynamically evolved. In fact, all of them belong to the oldest groups of Family II or Family III in the
dynamical classification proposed by \citet{Ferraro12}. Indeed, NGC 6397 and M30 are post-core collapsed clusters and
NGC 6752 is suspected to be in the post-core collapse bounce \citep{Ferraro03}.

\subsection{Chemical abundances}

\begin{figure}
\plotone{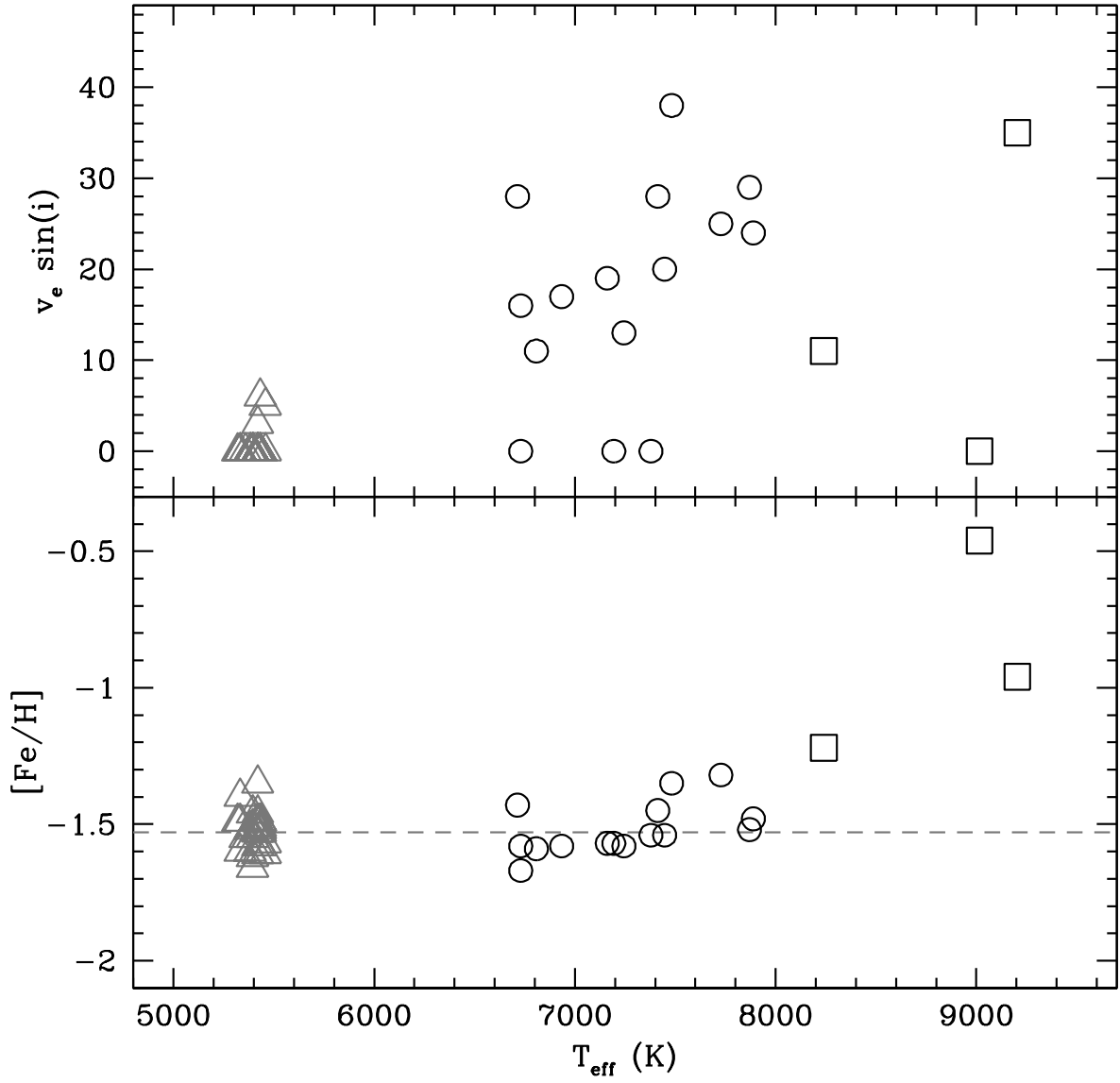}
\caption{Upper panel: rotational velocities for RGB stars (grey triangles) and BSSs (circles and squares) as a function of temperatures.
Lower panel: iron abundances for RGB stars and BSSs. The dashed horizontal line defines the mean Fe abundance of RGB stars.
Squares are BSSs affected by radiative levitation.}
\label{fe}
\end{figure}

\begin{figure}
\plotone{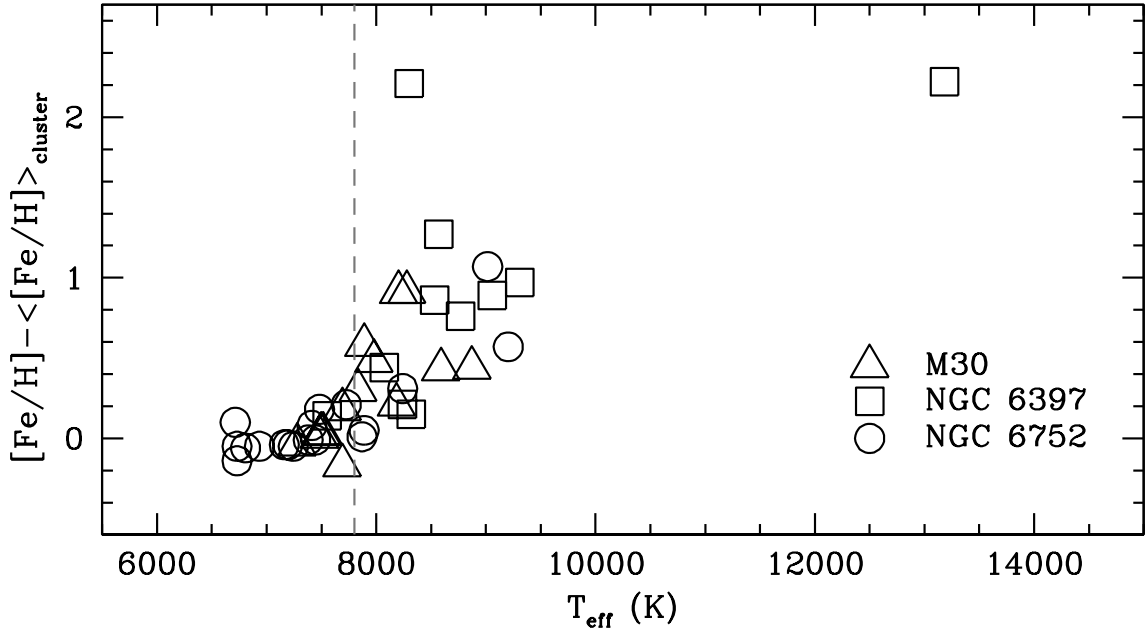}
\caption{[Fe/H] ratios (scaled to the mean metallicity of the clusters) as a function of temperature for BSSs in M30 (triangles, \citealt{Lovisi13}),
NGC 6397 (squares, \citealt{Lovisi12}) and NGC 6752 (circles, this work).
The dashed vertical line marks the temperature threshold for the onset of radiative levitation (T$_{eff}$=7800K).}
\label{lev}
\end{figure}

{\bf Iron --} As shown in Sect. \ref{rad}, Fe abundances for the observed BSSs help us to discriminate possible field stars. The [Fe/H] ratios for 
almost all BSSs agree very well with that derived for RGB stars and with previous results by other authors,
excluding that these BSSs belong to the field. The only exceptions are BSS1 ([Fe/H]$=-$1.22), BSS2 
([Fe/H]$=-$0.96) and BSS10 ([Fe/H]$=-$0.46). Nevertheless, from Fig. \ref{fe} it is clear that these stars are also 
the hottest ones in our sample, with T$_{eff}>$8000 K. \citet{Lovisi12} and \citet{Lovisi13} observed the same behavior
(the hottest BSSs were also the most metallic ones) in several BSSs of NGC 6397 and M30. 
These results have been interpreted in terms of the occurrence of radiative levitation in the shallow 
convective envelopes of BSSs. Radiative levitation is a diffusion process that occurs when radiative acceleration exceeds 
the gravitational one \citep{Michaud1983}: therefore, many chemical elements are pushed toward the stellar surface 
and the original chemical composition is altered (in particular, the iron-peak elements abundances are enhanced
with respect to the original chemical composition). In this framework, 
we conclude that also BSS1, BSS2 and BSS10 in NGC 6752 suffer from radiative levitation.
Fig. \ref{lev} shows the [Fe/H] ratio for all the BSS samples in NGC 6397, M30 and NGC 6752 as a function
of temperature. In order to compare in a meaningful way the iron abundances for BSSs in clusters with different metallicity,
BSS [Fe/H] ratios have been scaled to the mean cluster metallicities ([Fe/H]$=-$2.12 for NGC 6397, \citealt{Lovisi12} , 
[Fe/H]$=-$2.28 for M30, \citealt{Lovisi13} and [Fe/H]$=-$1.53 for NGC 6752, in this work. A clear trend with temperature exists and an abrupt 
increase in the [Fe/H] ratio is observed for BSSs hotter than 7800 K (dashed grey line). Independently from the cluster 
metallicity, the BSS iron abundance systematically increases with T$_{eff}$ up to more than 2 dex. 

{\bf Carbon and Oxygen --} Carbon and oxygen surface abundances can be used as tracers of mass transfer processes where
the donor star is peeled down to regions where the CNO-cycle occurred \citep{Ferraro06, Mucciarelli13b}.
In particular C and O depletion are expected for MT-BSSs \citep{Sarna96} whereas no chemical
anomaly should appear on the COL-BSSs surface \citep{Lombardi95}. The first observational confirmation has been provided by \citet{Ferraro06} 
that discovered 6 out 42 CO-depleted BSSs in 47 Tucanae. Moreover, \citet{Lovisi13} identified 4 BSSs with O
upper limits incompatible with the cluster O abundance in the GC M30. These stars lie along the BSS red sequence 
that has been interpreted by \citet{Ferraro09} as the one formed by MT-BSSs, at odds with the blue one formed by COL-BSSs. 
On the contrary, no evidence of CO-depletion has been found in M4 \citep{Lovisi10} whereas in the case of NGC 6397 the occurrence
of radiative levitation on the surface of almost all the BSSs in the sample, prevents the interpretation of O abundances in terms of 
BSS formation mechanisms \citep{Lovisi12}.

C and O abundances for the BSSs not affected by radiative levitation in NGC 6752 are in agreement with those derived
for dwarf stars \citep{Carretta05}. No CO-depleted BSSs have been identified in the studied sample. As suggested by 
\citet{Ferraro06} the CO-depletion could be a transient phenomenon and ''evolved" MT-BSSs might have restored their original 
chemical abundances.
According to the percentage of CO-depleted BSSs in 47 Tucanae (14\%), we could have expected 1-2 CO-depleted stars in NGC 6752. The lack of chemical
anomalies in the studied sample might suggest that all the (investigated) BSSs in NGC 6752 may derive from stellar collisions, 
for which no chemical anomalies are expected. This features is again in agreement with the particular dynamical state proposed for NGC 6752
by \citet{Ferraro03}.

\acknowledgements
{This research is part of the project COSMIC-LAB (www.cosmic-lab.eu) 
funded by the European Research Council
(under contract ERC-2010-AdG-267675). We thank the anonymous referee for his/her valuable suggestions.}

\begin{sidewaystable}
\scriptsize 
\centering
\medskip
\rotatebox{90}{}
\setlength\tabcolsep{4pt}
\begin{tabular}{ccccccccccccc}
\hline
\hline
ID & RA & DEC & V & I & T$_{eff}$ & $\log(g)$ & RV & v$_{e}$ $\sin(i)$ & [Fe~II/H] &  [O/H] & [C/H] & Notes\\ 
    & (degrees) & (degrees) & & & (K)&  & ($\kms$) & ($\kms$) & & & &\\
\hline
BSS1 &  287.7410103 &  -59.9633904 &  15.49 &  15.27 &  8241 &  4.0 & -22.6 $\pm$ 0.1 &11 $\pm$ 1 & -1.22 $\pm$ 0.09 & -1.71 $\pm$ 0.17 & -1.69 $\pm$ 0.14 & PM\\	 
BSS2 &  287.7168646 &  -59.9846242 &  15.62 &  15.48 &  9204 &  4.3 & -24.7 $\pm$ 0.8 &35 $\pm$ 1 & -0.96 $\pm$ 0.05 & -2.31 $\pm$ 0.03 & -2.30 $\pm$ 0.06 &\\	  
BSS3 &  287.7039030 &  -59.9910308 &  16.45 &  15.88 &  6808 &  4.0 & -19.5 $\pm$ 1.5 &11 $\pm$ 2 & -1.59 $\pm$ 0.09 & -1.65 $\pm$ 0.11 & -1.45 $\pm$ 0.07 &\\	  
BSS4 &  287.6827377 &  -59.9749054 &  16.59 &  16.15 &  7447 &  4.2 & -23.7 $\pm$ 2.2 &20 $\pm$ 2 & -1.54 $\pm$ 0.08 & -1.35 $\pm$ 0.08 & -1.65 $\pm$ 0.10 &\\	  
BSS5 &  287.6897573 &  -59.9918788 &  16.71 &  16.11 &  6730 &  4.0 & -29.5 $\pm$ 1.1 & 0 $\pm$ 1 & -1.67 $\pm$ 0.17 & -1.48 $\pm$ 0.15 & -1.27 $\pm$ 0.08 & PM\\	  
BSS6 &  287.6872156 &  -60.0099633 &  16.70 &  16.23 &  7161 &  4.1 & -27.2 $\pm$ 0.7 &19 $\pm$ 2 & -1.57 $\pm$ 0.12 &   --	&    -- & $\gamma$Dor\\	  
BSS7 &  287.7208140 &  -59.9780412 &  16.74 &  16.13 &  6730 &  4.0 & -22.7 $\pm$ 0.9 &16 $\pm$ 5 & -1.58 $\pm$ 0.12 & -1.47 $\pm$ 0.14 & -1.43 $\pm$ 0.07 &\\	  
BSS8 &  287.7141443 &  -59.9718582 &  16.72 &  16.38 &  7727 &  4.4 & -23.6 $\pm$ 1.3 &25 $\pm$ 5 & -1.32 $\pm$ 0.13 & -1.19 $\pm$ 0.08 & -1.75 $\pm$ 0.06 &\\	  
BSS9 &  287.6997850 &  -59.9785210 &  17.11 &  16.63 &  7244 &  4.3 & -33.3 $\pm$ 0.2 &13 $\pm$ 7 & -1.58 $\pm$ 0.10 & -1.82 $\pm$ 0.07 & -1.99 $\pm$ 0.06 &\\	  
BSS10 &  287.7669498 &  -59.9853719 &  15.33 &  15.19 &  9016 &  4.1 & -24.6 $\pm$ 0.7 & 0 $\pm$ 1 & -0.46 $\pm$ 0.08 & -2.68 $\pm$ 0.06 & -2.06 $\pm$ 0.06 & PM, EHB\\	  
BSS11 &  287.7252708 &  -60.0033895 &  15.74 &  15.26 &  7482 &  3.9 & -17.3 $\pm$ 4.4 &38 $\pm$ 1 & -1.35 $\pm$ 0.14 & -1.34 $\pm$ 0.08 & -1.62 $\pm$ 0.10 & PM, SX-Phe\\	  
BSS12 &  287.7294365 &  -59.9826813 &  16.24 &  15.91 &  7889 &  4.2 & -27.1 $\pm$ 1.3 &24 $\pm$ 4 & -1.48 $\pm$ 0.11 & -1.87 $\pm$ 0.06 & -2.03 $\pm$ 0.04 &\\	  
BSS13 &  287.7299968 &  -59.9810128 &  16.65 &  16.06 &  6714 &  4.0 & -14.1 $\pm$ 1.7 &28 $\pm$ 2 & -1.43 $\pm$ 0.14 &  -- &	      --  & \\	  
BSS14 &  287.7546346 &  -59.9891338 &  16.63 &  16.19 &  7379 &  4.2 & -35.0 $\pm$ 0.4 & 0 $\pm$ 2 & -1.54 $\pm$ 0.10 & -1.28 $\pm$ 0.08 & -1.41 $\pm$ 0.10 & PM\\  
\hline
ID & RA & DEC & B & R & T$_{eff}$ & $\log(g)$ & RV & v$_{e}$ $\sin(i)$ & [Fe~II/H] &  [O/H] & [C/H] & Notes\\ 
    & (degrees) & (degrees) & & & (K)&  & ($\kms$) & ($\kms$) & & & &\\
\hline
BSS15 & 287.7851741 & -60.0189240 &  16.17 & 15.68 & 7194 & 3.9 & -23.4 $\pm$ 0.4 & 0  $\pm$ 1 & -1.57 $\pm$ 0.08 & -1.23 $\pm$ 0.17 & -1.70 $\pm$0.17 & PM\\	  
BSS16 & 287.7918025 & -59.9806129 &  16.83 & 16.31 & 6934 & 4.1 & -35.5 $\pm$ 0.1 & 17 $\pm$ 4 & -1.58 $\pm$ 0.16 & -1.40 $\pm$ 0.09 & -1.86 $\pm$0.06 & PM\\	  
BSS17 & 287.6424824 & -59.9491533 &  16.21 & 15.83 & 7413 & 4.0 & -23.4 $\pm$ 3.6 & 28 $\pm$ 2 & -1.45 $\pm$ 0.07 & -1.09 $\pm$ 0.13 & -1.81 $\pm$0.08 & PM, SX-Phe\\	  
BSS18 & 287.5595535 & -60.0844359 &  16.35 & 16.04 & 7870 & 4.2 & -32.8 $\pm$ 0.1 & 29 $\pm$ 5 & -1.52 $\pm$ 0.17 & -1.0 1$\pm$ 0.20 & -1.33 $\pm$0.08 &\\	  
\hline
\end{tabular}
\caption{Identification numbers, coordinates, magnitudes, effective temperatures, surface gravities, radial and rotational velocities, 
Fe, O and C abundances of the BSSs in the ACS (on the top) and WFI (on the bottom) samples. PM indicates BSSs with proper motion by 
\citet{Zloczewski12}; $\gamma$Dor, EHB and SX-Phe are $\gamma$Dor pulsator stars, Extreme Horizontal Branch stars and SX-Phoenicis 
stars according to \citet{Kaluzny09}.}
\label{bss} 
\end{sidewaystable}

\begin{sidewaystable}
\scriptsize
\centering
\medskip
\rotatebox{90}{}
\setlength\tabcolsep{4pt}
\begin{tabular}{cccccccccc}
\hline
\hline
ID & RA & DEC & B & R & T$_{eff}$ & $\log(g)$ & RV & v$_{e}$ $\sin(i)$ & [Fe~I/H]\\ 
    & (degrees) & (degrees) & & & (K)&  & ($\kms$) & ($\kms$) &\\
\hline
RGB1 &  287.8801804 &  -60.0657549 &  15.71 &  14.75 &  5321 &   2.9 &  -31.0 $\pm$ 0.4 &  0 $\pm$ 1 & -1.49 $\pm$ 0.19\\  
RGB2 &  287.8762748 &  -60.0447351 &  16.34 &  15.41 &  5420 &   3.2 &  -23.9 $\pm$ 0.4 &  0 $\pm$ 1 & -1.58 $\pm$ 0.18\\
RGB3 &  287.8683233 &  -60.0417224 &  16.17 &  15.22 &  5395 &   3.2 &  -30.0 $\pm$ 0.4 &  0 $\pm$ 1 & -1.66 $\pm$ 0.17\\  
RGB4 &  287.8715668 &  -60.0333950 &  16.16 &  15.22 &  5395 &   3.2 &  -23.9 $\pm$ 0.3 &  0 $\pm$ 1 & -1.55 $\pm$ 0.18\\  
RGB5 &  287.9149051 &  -59.9760061 &  16.35 &  15.41 &  5420 &   3.2 &  -26.4 $\pm$ 0.4 &  3 $\pm$ 4 & -1.61 $\pm$ 0.21\\  
RGB6 &  287.9489041 &  -59.9747585 &  16.07 &  15.13 &  5383 &   3.1 &  -23.9 $\pm$ 0.4 &  0 $\pm$ 1 & -1.60 $\pm$ 0.18\\  
RGB7 &  287.9380203 &  -60.0425142 &  16.03 &  15.09 &  5383 &   3.1 &  -20.1 $\pm$ 0.3 &  0 $\pm$ 1 & -1.52 $\pm$ 0.21\\  
RGB8 &  287.6115858 &  -60.1160018 &  16.38 &  15.44 &  5433 &   3.3 &  -29.0 $\pm$ 0.3 &  0 $\pm$ 1 & -1.52 $\pm$ 0.24\\  
RGB9 &  287.7844635 &  -60.0714674 &  16.40 &  15.47 &  5433 &   3.3 &  -22.7 $\pm$ 0.3 &  0 $\pm$ 1 & -1.49 $\pm$ 0.21\\  
RGB10 &  287.5961780 &  -60.0743004 &  16.23 &  15.28 &  5395 &   3.2 &  -26.3 $\pm$ 0.3 &  0 $\pm$ 1 & -1.46 $\pm$ 0.22\\  
RGB11 &  287.7903486 &  -60.0508958 &  15.77 &  14.79 &  5333 &   3.0 &  -32.3 $\pm$ 0.4 &  0 $\pm$ 1 & -1.60 $\pm$ 0.22\\  
RGB12 &  287.7629329 &  -60.0411150 &  16.20 &  15.27 &  5395 &   3.2 &  -28.9 $\pm$ 0.4 &  0 $\pm$ 1 & -1.62 $\pm$ 0.20\\  
RGB13 &  287.7711636 &  -60.0293473 &  16.35 &  15.42 &  5433 &   3.3 &  -18.3 $\pm$ 0.5 &  6 $\pm$ 6 & -1.58 $\pm$ 0.17\\  
RGB14 &  287.8302739 &  -60.0643595 &  15.87 &  14.92 &  5358 &   3.1 &  -26.8 $\pm$ 0.3 &  0 $\pm$ 1 & -1.55 $\pm$ 0.21\\  
RGB15 &  287.7805513 &  -60.0460663 &  16.35 &  15.41 &  5420 &   3.2 &  -33.4 $\pm$ 0.4 &  0 $\pm$ 1 & -1.53 $\pm$ 0.22\\  
RGB16 &  287.8110505 &  -60.0122533 &  16.47 &  15.54 &  5458 &   3.3 &  -36.5 $\pm$ 0.4 &  5 $\pm$ 4 & -1.61 $\pm$ 0.20\\  
RGB17 &  287.5672807 &  -60.0886491 &  16.42 &  15.51 &  5433 &   3.3 &  -24.9 $\pm$ 0.3 &  0 $\pm$ 1 & -1.49 $\pm$ 0.20\\  
RGB18 &  287.4885332 &  -60.0152793 &  16.25 &  15.31 &  5420 &   3.2 &  -26.4 $\pm$ 0.4 &  0 $\pm$ 1 & -1.46 $\pm$ 0.19\\  
RGB19 &  287.4794481 &  -60.0369896 &  16.42 &  15.47 &  5433 &   3.3 &  -24.2 $\pm$ 0.5 &  0 $\pm$ 1 & -1.51 $\pm$ 0.22\\  
RGB20 &  287.4514453 &  -60.2169573 &  15.86 &  14.88 &  5333 &   3.0 &  -23.5 $\pm$ 0.3 &  0 $\pm$ 1 & -1.40 $\pm$ 0.22\\  
RGB21 &  287.5688340 &  -60.1870682 &  16.32 &  15.38 &  5420 &   3.2 &  -27.9 $\pm$ 0.3 &  0 $\pm$ 1 & -1.35 $\pm$ 0.23\\  
RGB22 &  287.6634437 &  -60.1823294 &  16.29 &  15.36 &  5420 &   3.2 &  -24.7 $\pm$ 0.4 &  0 $\pm$ 1 & -1.49 $\pm$ 0.21\\  
RGB23 &  287.6113178 &  -60.1784682 &  16.39 &  15.48 &  5433 &   3.3 &  -30.0 $\pm$ 0.4 &  0 $\pm$ 1 & -1.53 $\pm$ 0.23\\  
RGB24 &  287.7655149 &  -60.1501678 &  15.82 &  14.86 &  5333 &   3.0 &  -27.8 $\pm$ 0.3 &  0 $\pm$ 1 & -1.49 $\pm$ 0.19\\  
RGB25 &  287.7063369 &  -60.1300567 &  16.42 &  15.54 &  5458 &   3.3 &  -23.3 $\pm$ 0.3 &  0 $\pm$ 1 & -1.57 $\pm$ 0.26\\  
RGB26 &  287.8546812 &  -60.1840216 &  16.13 &  15.19 &  5395 &   3.2 &  -29.5 $\pm$ 0.3 &  0 $\pm$ 1 & -1.51 $\pm$ 0.24\\  
\hline
\end{tabular}
\caption{Identification numbers, coordinates, magnitudes, effective temperatures, surface gravities, radial and rotational velocities and 
Fe abundance of the RGB sample.}
\label{rgb} 
\end{sidewaystable}

\end{document}